\documentclass[Afour,sageh,times]{sagej}
\usepackage{amsmath,amssymb,amsfonts}
\usepackage{algorithmic}
\usepackage[colorlinks,,bookmarksopen,bookmarksnumbered,citecolor=red,urlcolor=red]{hyperref}
\usepackage{graphicx}
\usepackage{textcomp}
\usepackage{xcolor}
\usepackage{booktabs}
\usepackage{tabularx}
\usepackage{algorithm}
\usepackage{algorithmic}
\usepackage{natbib}
\usepackage{moreverb,url}
\usepackage[switch]{lineno}

\newcommand{\revise}[1]{#1}

\def\BibTeX{{\rm B\kern-.05em{\sc i\kern-.025em b}\kern-.08em
    T\kern-.1667em\lower.7ex\hbox{E}\kern-.125emX}}
\begin{document}

\def\volumeyear{2023}
\runninghead{Underwood et al}

\title{Black-Box Statistical Prediction of Lossy Compression Ratios for Scientific Data}
\begin{abstract}
Lossy compressors are increasingly adopted in scientific research, tackling volumes of data from experiments or parallel numerical simulations and facilitating data storage and movement. In contrast with the notion of entropy in lossless compression,  no theoretical or data-based quantification of lossy compressibility exists for scientific data. Users rely on trial and error to assess lossy compression performance. As a strong data-driven effort toward quantifying lossy compressibility of scientific datasets, we provide a statistical framework to predict compression ratios of lossy compressors. Our method is a two-step framework where (i) compressor-agnostic predictors are computed and (ii) statistical prediction models relying on these predictors are trained on observed compression ratios. Proposed predictors exploit spatial correlations and notions of entropy and lossyness via the quantized entropy. We study 8+ compressors on 6 scientific datasets and achieve a median percentage prediction error less than 12\%, which is substantially smaller than that of other methods while achieving at least a $8.8\times$ speedup for searching for a specific compression ratio and $7.8\times$ speedup for determining the best compressor out of a collection.  
\end{abstract}

\author{Robert Underwood\affilnum{1}, Julie Bessac\affilnum{1}, David Krasowska\affilnum{2}, Jon C. Calhoun \affilnum{2}, Sheng Di \affilnum{1}, and Franck Cappello \affilnum{1}}
\affiliation{\affilnum{1}Argonne National Laboratory, USA \affilnum{2} Clemson University, USA}
\corrauth{Robert Underwood, Argonne National Laboratory, Lemont, IL, USA}
\email{runderwood@anl.gov}

\maketitle


\section{Introduction}

The ever-increasing execution scale of high-performance parallel computing applications and advanced scientific facilities, producing extremely large volumes of data, presents challenges in scientific data storage and transfer.  The upcoming Linac Coherent Light Source II-HE, for example, will acquire data at a rate of 1 TB/s \cite{LCLS-HE} so that data cannot be stored without parallel lossy data reduction techniques. Lossless compression suffers from very low compression ratios (CRs) for these datasets, but  error-bounded lossy compression has been  effective in significantly reducing scientific data size with a strict control of data distortion. 
Significantly reducing the data size without sacrificing data integrity is a concerning research problem for many of today's scientific projects run on large parallel machines, especially because efficiently storing and transferring data are key for post hoc data analysis and management. Thus, many of today's scientific data formats, including NetCDF and HDF5,  support data reduction by calling various third-party data compression libraries.\footnote{\url{portal.hdfgroup.org/display/HDF5/HDF5+Dynamically+Loaded+Filters}} 

\begin{figure*}
    \centering
    \includegraphics[width=\textwidth]{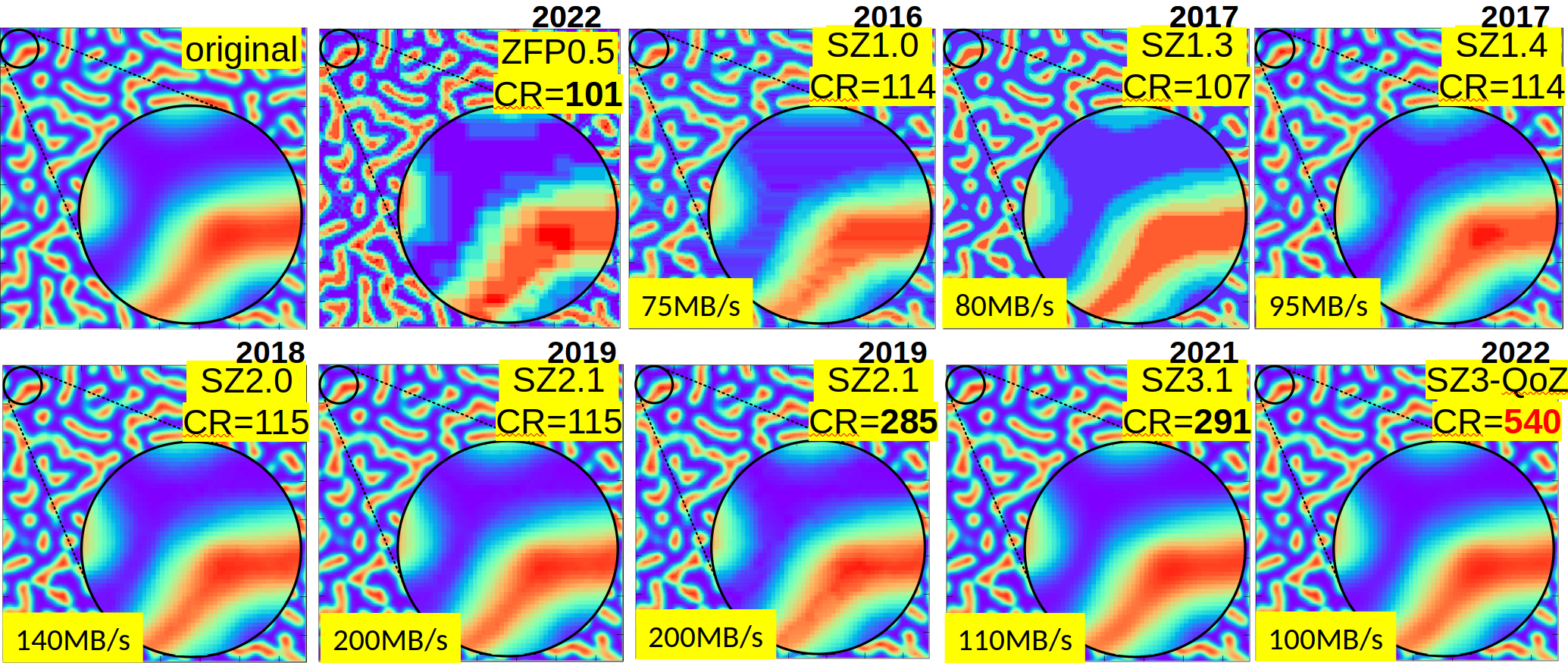}
    \caption{Visualization of Miranda - density data for SZ’s different versions (EB: value range relative error $10^{-2}$), Performance on single core CPU (Intel Broadwell). Versions: ZFP 0.5 \cite{lindstromFixedRateCompressedFloatingPoint2014}, SZ 1.0 \cite{diFastErrorBoundedLossy2016} SZ 1.3-4 \cite{taoSignificantlyImprovingLossy2017}, SZ2 \cite{liangErrorControlledLossyCompression2018}, SZ3 \cite{zhaoOptimizingErrorBoundedLossy2021}, SZ3-QoZ \cite{liuDynamicQualityMetric2022}}
    \label{fig:improvement}
\end{figure*}

Substantial progress has been made in the design of lossy compressors to improve their performance, enhance the quality assessment methodology, and expand the range of applications that can use lossy compression.
Lossy compressors can now achieve substantial compression ratios quickly while maintaining the scientific integrity of the data.
While images are not the only way to assess distortion, Figure~\ref{fig:improvement} gives an intuition of the improvement in both compression ratio (how much the data was reduced; larger is better) and quality for two of the leading lossy compressors for scientific data -- SZ and ZFP.
Compression is now used for many more use cases \cite{use-case} from classic use cases such as visualization, reducing storage footprint, and reducing I/O time, and now includes many new use cases reducing streaming intensity, lossy checkpoint/restart, avoiding re-computation with lossy caching, running larger simulations by reducing memory footprint, accelerating CPU $\leftrightarrow$ GPU transfers, reducing deep neural networks  model size, and accelerating training of deep neural networks. 

There have also been several advances in the development of methodologies to assess the quality of lossy compressors including the development of standard datasets with SDRBench \cite{zhaoSdrbench2020}, a consistent interface with LibPressio \cite{underwoodProductivePerformantGeneric2021}, and quality assessment tools including Z-Checker \cite{taoZcheckerFrameworkAssessing2019} and Foresight \cite{grossetForesightAnalysisThat2020}.
We expect the future will empower and demand many new uses of lossy compression which will require even further innovation in the design of lossy compressors.

However, with over half a decade of consistent improvements, an important question now arises: what is the limit of the compressibility of scientific data? This question is important, at least for two reasons: (i) researchers in lossy compression algorithms need to know if further progress is still possible. In the lossless context, the Shannon entropy and source coding theorem establish a theoretical lower bound of the code rate (average number of bits per symbol). Lossless coding algorithms were compared to this lower bound, which allowed the demonstration that arithmetic coding is nearly optimal. (ii) users of lossy compression need to know what compression ratio is achievable for their datasets, considering a quality tolerance. For lossy compression of scientific data, the community lacks a roof-line model relating the largest achievable compression ratios to different users' quality tolerances.
While establishing theoretical bounds of lossy compressibility for scientific datasets is currently a hard open research problem, a useful step in this direction is to produce models of compression ratio that are fast and highly accurate for many compressors. Such models would respond to users' need to estimate the compressibility (or expected compression ratio) of their datasets for several compressors before actually performing the compression operation. We focus on two existing typical use cases in parallel computing. (\textbf{use case 1}) Enabling fast, automated configuration of a single compressor (that will be run in parallel on many nodes)  to maximize the quality that will fit in available storage \cite{underwoodFRaZGenericHighFidelity2020}. Reducing the storage footprint needed by parallel applications such as cosmology simulations \cite{hacc}, climate simulations (e.g. the community earth system model (CESM)), and X-ray crystallography  \cite{Chuck1} that dump vast amounts of data to fixed-size external devices.
(\textbf{use case 2}) Enabling quickly choosing among several compressors with the highest CR at runtime in order to minimize data size \cite{taoOptimizingLossyCompression2019}.

However, existing lossy compression models are either too slow to be used in applications or too inaccurate to be effective.
Accurately predicting lossy compression ratios for scientific datasets is  challenging because of two key factors. (1) The lossless Shannon entropy alone cannot be used directly on scientific datasets to estimate the compressibility because they often exhibit a high level of autocorrelation that lossy compression techniques can leverage. (2) Lossy compression removes information from datasets, leading to changes in the symbol distribution, hence the need to quantify lossyness as part of a compressibility measure.

To address this gap, we present an efficient black-box statistical lossy compression ratio prediction method for 2D and 3D scientific datasets. 
Our method requires two steps: (1) we perform \textit{compressor-agnostic} statistical analysis on datasets, and (2) we train a statistical prediction model from resulting statistics of Step (1) and observed CRs from existing compressors. 
This is the first formulation of a generic (compressor-free) statistical prediction model that  uses training only from observed compression ratios for its specialization to a compressor. The generic model can specialize to any lossy compressor before prediction, as opposed to other prediction models that depend on knowledge of a compressor's design principles \cite{luUnderstandingModelingLossy2018,qinEstimatingLossyCompressibility2020,krasowska2021exploring} or need several trial runs during prediction \cite{underwoodFRaZGenericHighFidelity2020}. We note that all existing and  currently proposed methods require the use of compressors to generate compression ratios as training data.

\begin{figure}
    \centering
    \hspace{-3mm}
    \includegraphics[scale=.3]{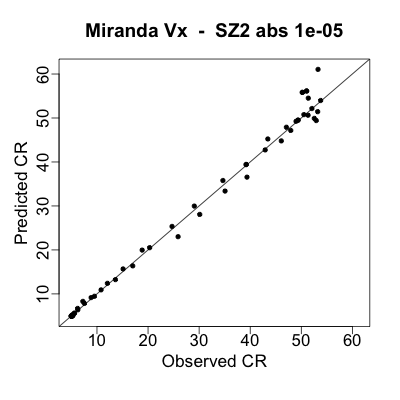}
    \hspace{-4mm}
     \includegraphics[scale=.3]{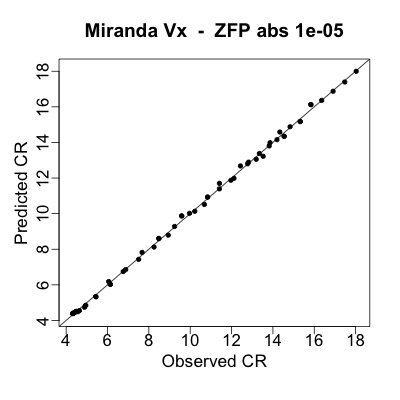} \\[-8pt]
      \includegraphics[scale=.3]{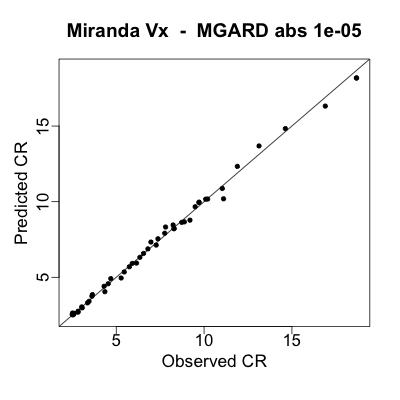}
    \hspace{-4mm}
     \includegraphics[scale=.3]{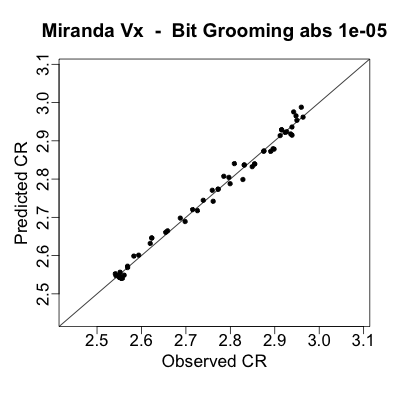}
     \\[-14pt]
    \caption{Scatter plots between true and out-of-sample predicted compression ratios from the proposed statistical models. Results are shown for the field velocity-x of Miranda for SZ2 (top left), ZFP (top right), MGARD (bottom left), and Bit Grooming (bottom right) with $10^{-5}$ absolute error bound. } 
    \label{fig:scatterplot_prediction}
\end{figure}

Our technical contributions are summarized as follows. 
(i) We derive data-based statistical predictors that exploit the spatial correlation of data via a singular value decomposition (SVD) and notions of entropy and lossyness via the quantized entropy, and we show their complementarity.
(ii) We study leading state-of-the-art lossy compressors covering different compression models with different principles. 
(iii) We carefully investigate various responses of our developed compression ratio predictions to the different data compressor-prediction schemes used in the state-of-the-art lossy compressor SZ, providing a more in-depth understanding of the compression quality of SZ2 compared with the prior works \cite{qinEstimatingLossyCompressibility2020, krasowska2021exploring} that studied SZ2.  SZ is widely used in the community, but its compression ratio remains challenging to predict because of its multiprediction-scheme approach. 
(iv) We evaluate our proposed statistical compression ratio prediction method on 6 real-world scientific simulation datasets and 4 kinds of synthetic samples (used to assess precisely the CR response to different levels of controlled correlations). We compare the prediction accuracy of our method with other competitive statistical prediction methods (block sampling and compressor-specific). Our method obtains a very high prediction accuracy, as exemplified in Fig.~\ref{fig:scatterplot_prediction} (median percentage of error of $2.8\%$ for SZ2~\cite{diFastErrorBoundedLossy2016}, $1.0\%$ for ZFP~\cite{lindstromFixedRateCompressedFloatingPoint2014}, $2.0\%$ for MGARD~\cite{ainsworthMultilevelTechniquesCompression2019}, and $0.3\%$ for Bit Grooming~\cite{zenderBitGroomingStatistically2016}) across a wide range of compression ratios from below $3$ to above $50$.
(v) We demonstrate a speedup for the two use cases above compared to traditional procedures. 
%


\section{Background and related work} \label{sec:background}


\subsection{Lossy compressors for scientific data}

Lossy compressors typically comprise (i) a decorrelation step that exploits  correlations present in the dataset to transform the data into a more compressible one for the following steps,
(ii) an approximation/quantization step that reduces the precision of the input data bringing the lossyness in the compression pipeline, and (iii) an encoding step that minimizes the number of bits used to represent the approximation step outcomes. 
Different lossy compressors leverage different decorrelation methods and thus have different responses to correlation structures of the data.
We describe several leading lossy compressors  in the chronological order of their introduction, paying  attention to how each leverages spatial and entropy information. 
Decorrelation steps in these compressors are based on transformation, prediction, or rounding, providing a wide range of compression behaviors in the study.

The ZFP compressor uses a near-orthogonal transform-based approach for decorrelation \cite{lindstromFixedRateCompressedFloatingPoint2014}.
This approach represents each $4^n$ block as a sum of possible spatial patterns, where $n$ is the number of dimensions of the considered dataset. 
Values in each block are converted to a shared fixed-point representation. After that, a near-orthogonal transform is applied to the fixed-point representation.
The transformed data is then encoded such that the least significant bits are truncated from each block to achieve a desired size or quality.
Because of this design, ZFP has knowledge of neighboring information up to three elements away in each direction.

The SZ series \cite{liangEfficientTransformationScheme2018, taoFixedPSNRLossyCompression2018, zhaoOptimizingErrorBoundedLossy2021, liangErrorControlledLossyCompression2018,liuDynamicQualityMetric2022} of error-bounded lossy compressors relies on prediction as a decorrelation principle. 
The SZ  compressors generally use a combination of a  Lorenzo-prediction scheme, which uses the immediate preceding values, and later a linear regression \cite{liangErrorControlledLossyCompression2018}, which fits a hyperplane in  each block of the  dataset.
If the prediction is close enough to preserve the error bound  in absolute, value-range relative \cite{liangErrorControlledLossyCompression2018}, or pointwise relative error bounds \cite{liangEfficientTransformationScheme2018} or to preserve a peak  signal-to-noise  ratio bound \cite{taoFixedPSNRLossyCompression2018}, then the value is stored lossily; otherwise it is stored losslessly. The coding uses the Huffman coding followed by an entropy encoding stage performed via ZStandard  or GZIP depending on the version and configuration.
For 2D data, SZ observes neighboring information up to 15 elements  with the regression-prediction scheme and up to 1 or 2 elements with the Lorenzo-prediction scheme, which increases the  capacity to leverage spatial features. 
If the value range of the data is less than the error bound, SZ collapses all values to a special header. We exclude this special case from our study  because it is easy to detect and does not leverage entropy or spatial relationships in the data. 

Bit Grooming, introduced in 2016 ~\cite{zenderBitGroomingStatistically2016}, operates by setting the most insignificant bits in the mantissa of IEEE floating-point values to either 0 or 1. The resulting data are more compactly compressed by sequence-recognizing compressors such as GZip or Zstandard. 
The Digit Rounding compressor operates similarly to Bit Grooming but instead rounds insignificant bits, as opposed to setting them to 0 or 1 \cite{delaunayEvaluationLosslessLossy2018}. 
These methods are largely unaware of spatial structures.

MGARD \cite{ainsworthMultilevelTechniquesCompression2018,ainsworthMultilevelTechniquesCompression2019,ainsworthMultilevelTechniquesCompression2019a} relies on mathematical multigrid methods and a hierarchical organization of the dataset. 
It decomposes  data into multilevel coefficients that represent recursively blocks until the block is represented within the allowed tolerance. 
These coefficients are then quantized and entropy coded with  Zlib and later Zstd. 
The coefficients represent regions of differing sizes and potentially the entire dataset. MGARD  captures multilevel effects that SZ and ZFP  may not, making it an important comparison for our paper.

SZ, ZFP, and MGARD have notions of absolute error bounds. 
The absolute error bound $\epsilon_{abs}$ is defined for any value $d_i$ in a dataset $D$ with corresponding decompressed value $\tilde{d}_i$ such that $| d - \tilde{d}_i | < \epsilon_{abs}$ for a specified error bound. 
Bit Grooming and Digit Rounding have alternate notions of precision, a form of pointwise relative error bound, which differs depending on the magnitude of the value.
By using the OptZConfig tool \cite{underwoodOptZConfigEfficientParallel}, one can automatically determine corresponding absolute error bounds accurately and efficiently.

We  consider only the TTHRESH compressor \cite{ballester-ripollTTHRESHTensorCompression2020} in Sect. \ref{sec:results3d} because it relies heavily on the higher-order SVD and hence  is the only relevant 3D+ context. It is also different in that it compresses the data as a whole, rather than  by blocks, thus allowing it to exploit long-range correlations.

\subsection{Existing methods for CR prediction} \label{sec:literature}

Several works in recent years have attempted to estimate CR of lossy compressors for UCs in parallel computing.
The vast majority of prior approaches are white-box approaches such as \cite{liangImprovingPerformanceData2019, luUnderstandingModelingLossy2018, qinEstimatingLossyCompressibility2020, liangSignificantlyImprovingLossy2019,taoOptimizingLossyCompression2019}. 
Techniques in \cite{qinEstimatingLossyCompressibility2020, luUnderstandingModelingLossy2018} operate by respectively  constructing a Gaussian distribution and a deep neural network  using internal statistics acquired from the compressors.  For example, Lu et al.~\cite{luUnderstandingModelingLossy2018} use  specific implementation details such as the number of nodes in the Huffman tree in order to estimate CR for SZ1.4, hence depending on the compressor-prediction performance and the entropy of a particular dataset. 
Moreover, this approach requires the execution of almost the entire compression pipeline, except writing the final output buffer of SZ1.4. 
The authors also assume that the number of Huffman nodes can be approximated by a Gaussian distribution.  
Such techniques may not generalize and require expert knowledge about the compressor to formulate. 
Furthermore, they are tied to the evolution of the compressor; for example, the new prediction procedures in SZ such as regression and interpolation were introduced after this work.

Techniques in \cite{liangImprovingPerformanceData2019, liangSignificantlyImprovingLossy2019, taoOptimizingLossyCompression2019} utilize a simpler model, albeit with less predictive power.
These techniques attempt to predict whether  SZ2 or ZFP has a better average bit rate (bits/symbol)  for a given set of blocks of data.
 Therefore, these techniques do not need to be accurate as long the predicted $1^{st}$ place ranking of compressors remain accurate.  
These prior works leverage sampling to reduce the volume of data to consider. 
Our work does not yet leverage sampling because of the possibility of the sensitivity to the block size;  we leave  this  for future work. 

Black-box approaches have been proposed to estimate compressor configurations that will result in a given CR  \cite{underwoodOptZConfigEfficientParallel,underwoodFRaZGenericHighFidelity2020} for SZ2, ZFP, and MGARD. 
These approaches leverage black-box techniques that construct a piecewise linear model of the domain to locate a specific CR by minimizing the anticipated error.
The techniques are relatively expensive, requiring many invocations of the compressors to get a good estimate of the CR for future points, thus limiting their usability.

Recently, a few works have investigated spatial and temporal statistical predictors of CR. 
In \cite{krasowska2021exploring}, global and local spatial correlation ranges that are estimated via a variogram are explored as candidate statistical predictors for CR. 
No CR prediction setup is proposed, however; and, as shown  in Fig.~\ref{fig:individual_predictor_miranda}, the spatial correlation on its own is insufficient to fully characterize 2D slices in terms of CR. 
In \cite{moon2022prediction} a prediction setup solely based on statistics of the data is proposed for a discrete cosine transform-based lossy compressor and applied to environmental time series.
The mean, variance, and skewness are considered,  as well as statistical properties of the time series, such as stationarity test outputs, time differences, and their spread. 
Several prediction techniques from ML/AI are compared. The study shows promising CR prediction accuracy; however, it is dedicated to time series and requires large amounts of data to train as they rely on neural network models. 

We will evaluate our approach in comparison to the leading methods that predict compression ratio in Section~\ref{sec:comparison}.

\subsection{Data} \label{sec:data}
In the following, we investigate  real-world scientific data from parallel applications as well as synthetic 2D-Gaussian samples with a known and controllable correlation structure. 
Gaussian samples provide a proof of concept, since their spatial correlation structures and intensities are explicitly known. We increase their complexity to mimic more realistic properties of real-world datasets and to generate challenging use cases.   

\subsubsection{Scientific datasets}
We explore 2D and 3D slices of 3D and 4D datasets.
We analyze several datasets from the Scientific Data Reduction Benchmark  \cite{zhaoSdrbench2020}, which have been produced by a variety of scientific simulations - many parallel. The effects of lossy compression on these datasets on these applications have been extensively investigated \cite{use-case}.  
We study in more detail a few fields (i.e., variables) from several datasets (typically a collection of several fields). 
In the following, we focus mostly on 2D slices from each field for various reasons: (a) this approach eases the visual inspection and intuition of spatial correlation and patterns, (b) the manipulation of statistical predictors defined in Sect.~\ref{sec:predictors} is easier in 2D, and (c)  slicing creates samples that are used to train statistical prediction models without using block-size dependent techniques as in \cite{qinEstimatingLossyCompressibility2020}. 
We slice 3D and 4D datasets along their slowest incrementing dimension to make more data available for training. This approach is not unreasonable since many communities (such as CESM) view their data slice by slice.

\begin{figure}
    \centering
    \includegraphics[scale=.37]{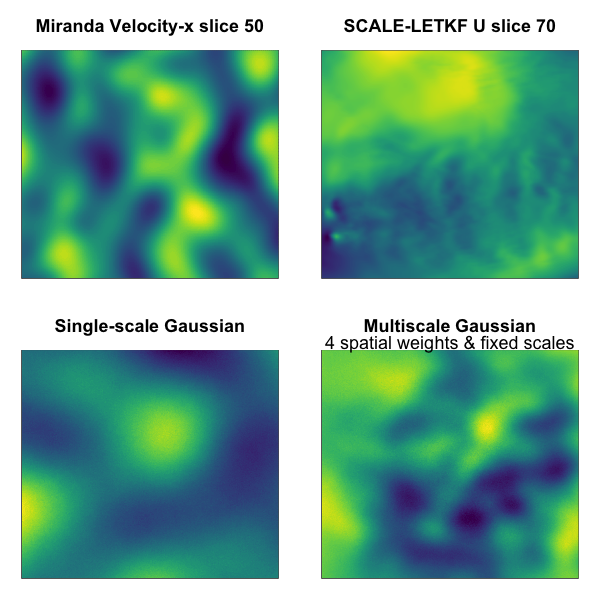} \\[-15pt]
    \caption{Top: 2D slices from Miranda velocity-x (left) and SCALE-LetKF U (right). 
    Bottom: 2 types of Gaussian samples: single-correlation sample (left) and multiscale samples spatial weights with fixed correlation ranges (right).} 
    \label{fig:gaussian_sample}
\end{figure}

 Miranda\footnote{\url{https://wci.llnl.gov/simulation/computer-codes/miranda}} is a radiation hydrodynamics code designed for large-eddy simulation of multicomponent flows with turbulent mixing. Each field is of dimension $256\!\times\!384\!\times384$. We reduce our focus to 2D slices $384\!\times\!384$ of its \textrm{velocity-x} (also denoted \textrm{Vx} below) and \textrm{density} (also denoted \textrm{De} below) fields that respectively represent fluid velocity along the x-axis and fluid density (fields of velocities along y- and z-axis, viscosity, pressure and diffusivity are also available). 
 We consider several climate simulation data from SDRBench, 
 in particular, data from the Community Earth System Model (CESM)\footnote{\url{http://www.cesm.ucar.edu}} atmospheric simulation code. 
 Most CESM fields are 2D ($1800\!\times\!3600$), and we focus on the cloud fraction, denoted CLOUD hereafter.  
Data from Hurricane-ISABEL\footnote{\url{http://sciviscontest-staging.ieeevis.org/2004/data.html}} is also studied.  
This dataset comes from simulations of Isabel, the strongest hurricane of the 2003 Atlantic hurricane season. 
Each field is a 3D array of $100\!\times\!500\!\times\!500$, and we consider $500\!\times\!500$ slices of the field \textrm{U} that is the east-west wind component. 
Additionally, the pressure and wind U fields of dimension $1200\!\times\!1200$  from the rainfall simulation  \revise{SCALE-LetKF}\footnote{\url{https://scale.riken.jp/scale-rm/}} are investigated. 
 Figure~\ref{fig:gaussian_sample} shows example slices of Miranda's \textrm{velocity-x} and \revise{SCALE-LetKF}'s U.
More illustrations of these datasets are available in \cite{zhaoSdrbench2020}. 
 We also investigate data coming from NYX\footnote{\url{https://amrex-astro.github.io/Nyx}}, a compressible cosmological hydrodynamics simulation code. Fields are represented by 3D arrays $512\!\times\!512\!\times\!512$. We focus on 2D slices of the \textrm{velocity-x} field of size $512\!\times\!512$ particle velocity along the x-axis.  
QMCPack\footnote{\url{qmcpack.org}} is an open source ab initio quantum Monte
Carlo package for analyzing the electronic structure of atoms,
molecules and solids. Each simulated orbital is  3D field ($96\!\times\!96\!\times\!115$) in single-precision. 
 See Table \ref{tab:datasets} for a summary, all data are dense, structured-grid, and IEEE 32 bit floating point. 
These datasets have various sizes of 2D slices that influence the compression's performance. However, we observe consistent results of compression ratio prediction for various slice sizes.

%
\begin{table}[]
    \centering
    \footnotesize
    \caption{Datasets}
    \label{tab:datasets}
    \begin{tabular}{lll}
    \toprule
         Name  & Dataset Size  & Domain  \\
     \midrule
          Miranda   & $384\times384\times256$  & Hydrodynamics \\
          Hurricane  & $500\times500\times100$  & Weather\\
          CESM & $3600\times1800\times1$  & Climate \\
          \revise{SCALE-LetKF} & $1200\times1200\times96$ &  Weather \\
          Nyx  & $512\times512\times512$ & Cosmology \\
          QMCPack & $96\times96\times115\times288$ &  Quantum Monte Carlo \\
          
    \bottomrule
    \end{tabular}
\end{table}



%

\subsubsection{Single- and multiscale Gaussian samples}\label{sec:gaussian}
To assess precisely the response of CR to different levels of controlled correlations and to generate 2D samples independent from one another (hence free from the slicing procedure), we generate 2D Gaussian samples $1028\!\times\!1028$. We simulate samples with a single correlation length and samples with multiple correlation ranges as weighted sums of single-correlation samples, similar to an SVD of spatial and spatiotemporal fields \cite{hannachi2007}, $\displaystyle X=\sum_{l=1}^{L}\omega_{l}U_{l}$ with $L\geq1$ and: \\
$\cdot$ $U_l$ is a 2D Gaussian sample with a squared-exponential correlation ${\boldsymbol {\Sigma }}(x_i,x_j) \!\! = \!\! \sigma^2 \exp( -|x_i - x_j|^2/a_l^2 )$, where the variance $\sigma^2$ is set to $1$, $a_l$ is the correlation range that is known and varied for the experiments, and $x_i$ are the spatial grid points of the 2D slices. \\
$\cdot$ The weights $(\omega_{l})_{l \in L}$ are specified in two ways: \begin{enumerate}
        \item $\omega_{l}$ is scalar and fixed at values in $[0.6,1.2]$.
        \item $\omega_{l}\!\!=\!\!\left(\det \nolimits ^{*}(2\pi \Omega )\right)^{-{\frac {1}{2}}}\,e^{-{\frac {1}{2}}(x - \mu )^{\!{\mathsf {T}}} \Omega^{+}(\mathbf {-1} -\mu )}$ is a 2D spatial Gaussian weight, taking values in $[0,1]$  with $x\!=\!(x_i, x_j)$ varied 2D-mean $\mu$ and fixed diagonal $\Omega$ to create various spatial patterns in the weights. 
    \end{enumerate}
$\cdot$  L=1, 3; when $L=3$,  
if the weights $\omega$ are scalar, $a_l$ are fixed or randomly drawn as a mixture of short, medium, and strong correlations. We restrict our focus to L=3 different scales for simplicity. 
This provides four types of Gaussian samples: (1) single-correlation samples,  multiscale correlation samples with (from simplest to most complex); (2) scalar weights and fixed correlation ranges; (3) spatial weights and fixed correlation ranges; and (4) spatial weights and random correlation ranges. The simplest and most complex are exemplified in Fig.~\ref{fig:gaussian_sample}.
Multiscale samples with scalar weights have a uniform contribution over the 2D slice of  each single-correlation sample, whereas samples generated with spatial weights exhibit more spatial heterogeneity as the spatially varying weights $\omega$ enforce different areas of the 2D slice to have different correlation strengths; see the bottom-right panel of Fig.~\ref{fig:gaussian_sample}.    

\subsection{Problem formulation and design overview} \label{sec:design}
We build a CR prediction framework (Sect.~\ref{sec:regression})  for the compressors described above, based on selected statistics of the data that serve as statistical predictors (Sect.~\ref{sec:predictors}).  
The CR is the ratio of the original data size to  the compressed data size. 
From the studied compressors, CRs are extracted for error bounds among $10^{-5}$, $10^{-4}$, $10^{-3}$, and $10^{-2}$ in absolute error mode. 
Most results are shown for a single error bound that  is prescribed by the user following user constraints and reconstruction quality or that provides realistic CRs for the data value range. 
CRs exhibit various ranges of value depending on the compressor, error bound, and data properties. 
Bit Grooming and Digit Rounding are run in absolute error bounds thanks to the correspondence provided by OptZConfig discussed above. 
To mimic most operational conditions, we focus on  CRs less than or equal to $100$. In practice, few users work with higher compression ratios (except for visualization, which is not a target UC of this study). Cappello et al.~\cite{use-case} report UCs with CRs around 150 and up to 1000 for data summary and visualization~\cite{biswas2020Sampling}. Nonetheless, our framework shows flexibility toward high-CR cases. For instance, running our method with CR up to 2000 leads to a difference in prediction accuracy of  1\% for a field of \revise{SCALE-LetKF} with CR greater than 100.  Similar tests for Miranda lead to an average difference of accuracy of  6.3\% for SZ, 4.3\% for ZFP, and 7.5\% for MGARD, indicating the robustness of our method to high-CR settings. 
Compressors are run with their default parameters (including default block size) and in absolute error mode. 
Software and compressor versions are the latest available on Spack with an extra repository \cite{underwoodRobertu94SpackPackages2020}. 
All experiments are run on a node with two 32-core Intel(R) Xeon(R) Gold 6148 CPU $@$ 2.40 GHz and 384 GB of RAM and an Nvidia v100 GPU.
The operating system is Linux CentOS 8 with compiler GCC 8.4.1.\footnote{For more details see the reproducibility code: \url{https://github.com/FTHPC/Correlation_Compressibility}.}

\enlargethispage{\baselineskip}
\section{Statistical prediction of compression ratios}  \label{sec:methods}
This section describes the identified statistical predictors based on spatial correlation and entropy and its quantized versions (Sect.~\ref{sec:predictors}). 
To predict CRs from selected statistical predictors, one needs to mathematically formalize the relationship between the CR and its predictors. Regression models are natural candidates since they express and approximate a response variable as a combination of identified predictors. 
In Sect.~\ref{sec:regression} we describe the linear and spline regressions that model the relationship between CR and its statistical predictors, along with the validation setting and metrics used to evaluate the quality of CR predictions (Sect.~\ref{sec:prediction}).

\subsection{Statistical predictors of compression ratios} \label{sec:predictors}
We describe statistical predictors that likely influence the decorrelation step \cite{use-case}  of the studied lossy compressors.  Correlation structures and patterns in the data are expected to influence the decorrelation step and hence the compressibility. 
Hereafter, the identified statistical predictors show a strong complementarity, providing a flexible set of predictors to CR. Rounding-based compressors do not necessarily have a decorrelation step; we will see that the entropy then plays an important role in these cases. 

\begin{figure}
    \centering
    \includegraphics[scale=.7]{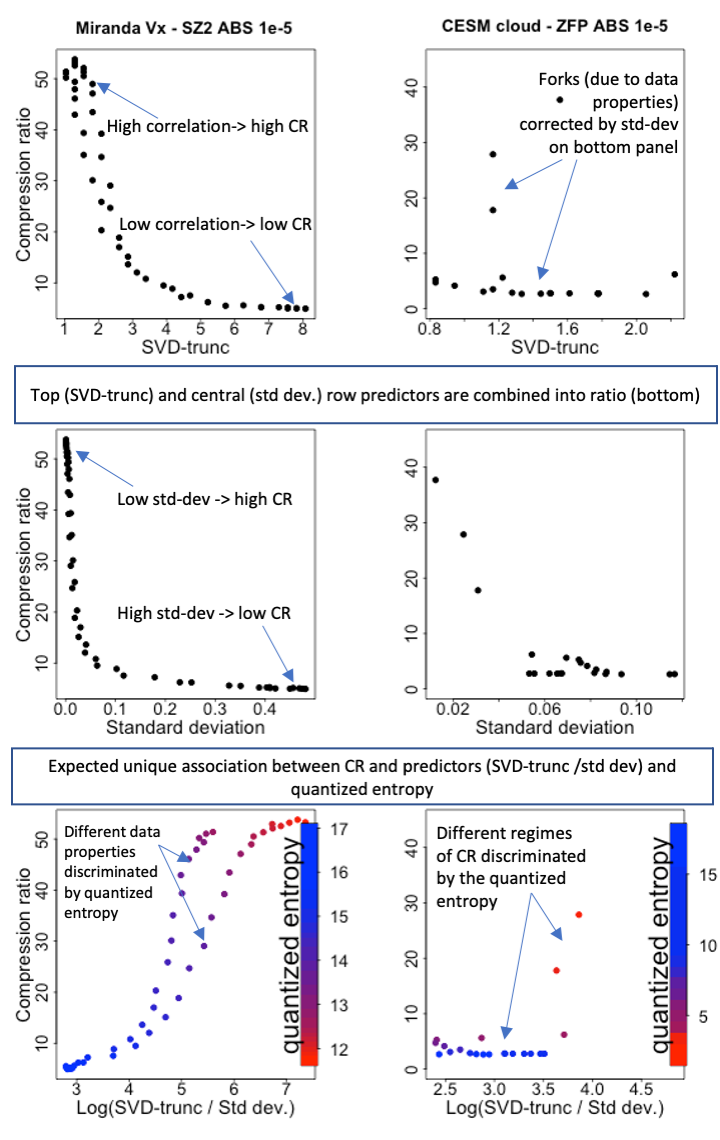}\\[-10pt]
    \caption{Relationship between compression ratios (y-axis) and their statistical predictors (x-axis). Statistical predictors are shown in the following order: SVD-truncation level (top x-axis), standard deviation (central x-axis), and their ratio $\frac{\mbox{SVD-trunc}}{\sigma}$ (bottomx-axis). Results shown for  Miranda velocity-x (left) and CESM cloud (right) fields. Each dot represents CR and statistics computed in a single 2D slice of each dataset.  
    The quantized entropy is  displayed through a color scale in the bottom panels.
    All results are for SZ2 (right) and ZFP (left) with absolute error bound $10^{-5}$ following user requirements.}
    \label{fig:individual_predictor_miranda}
\end{figure}

\subsubsection{Spatial correlation and SVD}
A few, recent studies show spatial correlation is intuitively an influential factor of transformation- and prediction-based compressors. 
In \cite{klower2021compressing}, the concept of bitwise real information (BIR) is introduced as the mutual information of bits in adjacent grid points.  
In particular, the stronger the association with neighboring bits, the greater the BIR. 
In \cite{krasowska2021exploring}, global and local measures of spatial correlation are introduced as explanatory variables of the CR. 
We next consider a proxy for spatial correlation based on  the SVD\footnote{SVD is linked to the Karuhnen--Loeve (KL) decomposition and its empirical version, the principal component analysis. 
KL orthogonally diagonalizes the covariance, hence provides a fully decorrelated representation in the Gaussian case, maximizing the coding gain \cite{gersho2012vector}. }  \cite{hannachi2007}  that reduces  computation times compared with estimating correlation range via a variogram \cite{matheron1963}.     
Specifically, the local variogram estimation on large slices of NYX (1200$\times$1200) from \cite{krasowska2021exploring} takes 17s  whereas the SVD takes 0.44s.

The SVD of a $m$$\times$$n$-matrix  $X$ provides the following decomposition: $X\!=\!U\Sigma V^{*}$ with  U an $m\!\times\!m$ complex unitary ($U U^{*}\!=\!I_{m}$) matrix (left singular vectors), $\Sigma$  an $m\!\times\!n$ rectangular diagonal matrix with non-negative real numbers on the diagonal (diagonal entries $\sigma _{i}=\Sigma _{ii}$ are the singular values), and V  an $n\!\times\!n$ complex unitary ($V V^{*}\!=\!I_{n}$) matrix (right singular vectors). 
We consider the percentage of singular values needed to recover $99\%$ of the total variance of $X$, which is proportional to $\sum_{i=1}^{\min(n,m)}\sigma_{i}^{2}$, when $X$ has mean-corrected columns. 
We use this truncation level as a proxy to spatial correlation range and denote it as ``svd-trunc'' in the following. 
The SVD-trunc is associated with the spatial correlation of a field. Small truncation levels indicate highly correlated fields for which few singular modes are needed to capture most of the variability In contrast, high truncation levels indicate that a high number of singular modes is required to capture most of the variability and hence a weakly correlated field. 
Analogous concepts for 3D datasets are discussed in the next paragraph, and associated results of CR predictions are presented in Sect.~\ref{sec:results3d}. 
Figure \ref{fig:individual_predictor_miranda} highlights almost one-to-one associations between CRs and SVD-truncation levels for both compressors and datasets. Highly correlated slices lead to high CRs and vice versa. However, different CR values can be associated with the same SVD-truncation level. Section~\ref{sec:stddev} and following discuss the resolution of this.

\subsubsection{3D-spatial correlation and high-order SVD}
The high-order SVD (HOSVD) is a type of Tucker decomposition \cite{tucker1963implications} for tensors and is seen as the extension of  SVD to a higher-order tensor context \cite{kolda2009tensor}.  In Sect. \ref{sec:results3d} we consider third-order (3D) real-valued tensors in $\mathcal{R}^{I\! \times\! \!J\! \times \!K}$. 
The Tucker decomposition provides a core tensor $G$ and matrices $A^{(i)}$ alongside each mode (dimension) $i$, for $i=1,...,N$ ($N=3$ in the third-order/3D case), which expresses a third-order tensor $X$ as $X=G \times_{1} A^{(1)} \times_{2} A^{(2)} \times_{3} A^{(3)}$, where $\times_{i}$ is the ith-mode product performing a matrix-like product on the ith-mode of the left hand-side tensor. 
The HOSVD of an $N$-order tensor $X$, the core tensor $G$, and matrices $A^{(i)}$ are calculated by 
unfolding dimension-$i$ to arrange the dimension-$i$ fibers (a fiber is a generalization of column to tensors) as columns into a 2D-matrix on which a traditional SVD is used.  
$A^{(i)}$ consists of the left singular vectors of this SVD. 
%
%
Kolda and Bader~\cite{kolda2009tensor} highlight the non-uniqueness in algorithms to choose the dimension of the core tensor $G$. 
Our method in Sect.~\ref{sec:predictors} operates by sorting and selecting singular values from each unfolded matrix such that the total contribution of the squared singular values exceeds $90\%$ per dimension.  


\subsubsection{Standard deviation}\label{sec:stddev}
In addition to accounting for the correlation strength between grid points, we consider the standard deviation $\sigma$ of each slice as an overall measure of the entire slice's variability and an indicator of its value range.
The central panels of Fig.~\ref{fig:individual_predictor_miranda} highlight the relationship between $\sigma$ and the CR, since slices with high variance are less compressible than slices with low variance. 
We note that $\sigma$ and the CR present a cleaner one-to-one association than the association between svd-trunc and the CR. 
 For some datasets, however, $\sigma$ is insufficient to fully describe the CR as a single predictor.  
The SVD-truncation level is then combined with the $\sigma$ via the ratio $\frac{\mbox{SVD-trunc}}{\sigma}$ in order to improve the one-to-one association between the CR and $\frac{\mbox{SVD-trunc}}{\sigma}$.  
In Fig.~\ref{fig:individual_predictor_miranda} the benefit of considering $\sigma$ is illustrated since the SVD-truncation level (top panels) does not discriminate all the CRs as they align vertically for similar values of SVD-truncation levels.
In contrast, on the bottom panels, the ratio $\frac{\mbox{SVD-trunc}}{\sigma}$ enables us to discriminate these points and suppress vertical alignments, hence improving the predictive power of the ratio $\frac{\mbox{SVD-trunc}}{\sigma}$ over the SVD-truncation level.  
In the following, we use the logarithm of $\frac{\mbox{SVD-trunc}}{\sigma}$ because $\sigma$ may be small. 

\subsubsection{Entropy} 
The entropy is viewed as an upper bound, associated with encoding, on the compression ratio for lossless compressors \cite{shannonMathematicalTheoryCommunication1948}.
It is defined as $H\left(D\right) = - \sum_{i \in D}\left(P\left(d_i\right) \log_2{P\left(d_i\right)}\right)$,  
where $P(d_i)$ is the probability of the symbol $d_i$ occurring in the dataset $D$.
It provides a good estimate of occurrence for compressibility purposes in a dataset, as it multiplies the relative frequency of a symbol by an idealized number of bits that could be used to represent the symbol.
A key limitation of entropy is that it does not account for  loss of information. 
The bottom panels of Fig.~\ref{fig:individual_predictor_miranda} illustrate the benefit of considering the entropy and its quantized version, as discussed below. 
The bottom left panel exhibits different intensity (via different colors) of the quantized entropy on the different forks of the scatterplot, indicating the complementarity of the ratio $\frac{\mbox{SVD-trunc}}{\sigma}$ and the quantized entropy. 
Similarly, the bottom left panel suggests a regime switch in the scatterplot for different intensity of  the quantized entropy. 
Variants of the entropy  \cite{claramunt2005spatialentropy, wang2018spatialentropy}  attempt to capture spatial patterns by representing the data topologically or as a graph and using notions of distance to scale the contributions of individual symbols in the encoding.
We leave extensions of our quantized entropy concept to future work. 

\subsubsection{Quantized statistics}
Next we apply quantization to the data prior to computing entropy. 
Quantization is a process that maps a continuous domain onto a discrete domain.
One of the most common forms of quantization is linear quantization, which is formulated as  $Q(d_i, \epsilon_{abs}) = \lfloor d_i/\epsilon_{abs} \rfloor * \epsilon_{abs}$, where $\lfloor x\rfloor$ is the floor function and $\epsilon_{abs}$ is the chosen number of subdivisions of the domain.  
We use this operation as a computationally inexpensive way to account for the maximum information loss after applying an absolute error bound, and we then compute the remaining entropy. 
Considering this lossyness allows for a clearer picture of how compressible the data is.
We define the quantized entropy as the entropy of the quantized data.  
In Fig.~\ref{fig:individual_predictor_miranda}, for both compressors (SZ on the left and ZFP on the right) and datasets, several values of CRs are associated with the same value of $\frac{\mbox{SVD-trunc}}{\sigma}$ (dots aligning vertically), indicating that $\frac{\mbox{SVD-trunc}}{\sigma}$ may not be sufficient on its own to fully characterize the CR. 
However, the corresponding quantized entropy has different intensity (represented by different colors) indicating that $\frac{\mbox{SVD-trunc}}{\sigma}$ and quantized entropy are complementary to characterize  CRs.   
For the studied compressors and datasets, similar trends are observed between the CR and the identified statistical predictors $\frac{\mbox{SVD-trunc}}{\sigma}$ and quantized entropy. 

\noindent\fbox{%
    \parbox{0.97\linewidth}{%
        \textbf{Key findings}: Statistical predictors $\frac{\mbox{SVD-trunc}}{\sigma}$ and quantized entropy together provide complementary explanatory power (Fig.~\ref{fig:individual_predictor_miranda}) for CR from studied compressors and  datasets.
    }%
}

\subsection{Linear and spline regressions} \label{sec:regression}
To predict CRs, we rely on regression models to model the relationship between the CR and its statistical predictors from Sect.~\ref{sec:predictors}.
For each compressor and each dataset field, regression models are fitted between the statistical predictors and associated compression ratios.  
We consider two regression models in this work:  contributions of the individual predictors from Sect.~\ref{sec:predictors} and their interaction as we observed their complementarity. 
We first consider a linear regression model because it provides the most intuitive setting with comprehensive parameters (further investigated in Sect.~\ref{sec:reg_coeff}):
\begin{eqnarray}\label{eq:linreg}
  \nonumber  \log(\mbox{CR}) &=& a + b\times\log(\mbox{q-ent}) +  c\times\log\left(\frac{\mbox{SVD-trunc}}{\sigma}\right) \\  
   \nonumber  && +  d\times\log(\mbox{q-ent})\times\log\left(\frac{\mbox{SVD-trunc}}{\sigma}\right)+\epsilon, \\
    && 
\end{eqnarray}  
where $\epsilon$ is a Gaussian random variable with mean $0$ and standard deviation $\sigma_{eps}$. 
Coefficients $a$, $b$, $c$, $d$, and $\sigma_{eps}$ are estimated by least-squares estimation with the {\rm R}-function  {\rm lm}. 
However, as relationships between the CR and its statistical predictors exhibit nonlinearity depending on the datasets (see Fig.~\ref{fig:individual_predictor_miranda}), a spline regression, often called generalized additive model or GAM, is also considered to account for more complex dependencies between the CR and its predictors:
\begin{eqnarray} \label{eq:splinereg}
  \nonumber  \log(\mbox{CR}) &=&  s(\log(\mbox{q-ent})) +  s\left(\log\left(\frac{\mbox{SVD-trunc}}{\sigma}\right)\right)  \\ 
\nonumber  && +  ti\left(\log(\mbox{q-ent}),\log\left(\frac{\mbox{SVD-trunc}}{\sigma}\right)\right)+\epsilon, \\
&& 
  \end{eqnarray} 
where $s$ is a cubic spline and $ti$ a tensor product spline that models the interaction between the two predictors. 
Spline regressions are typically used when relationships between the response variable and its predictors exhibit more complexity than linear or polynomial behaviors. Cubic splines are one of the most commonly used splines as they provide flexibility and regularity via piecewise third-order polynomials, while maintaining computational efficiency (coefficients computation reduces to a tridiagonal linear system). 
The number of knots to represent splines and tensor product splines is kept small (at 3) to prevent overfitting and because the number of data points is limited (between 30 and 199). 
The {\rm R}-package {\rm mgcv} is used to fit the spline regressions and perform prediction \cite{wood2017gam}. 
For both regression models, we consider the logarithm of CRs in order to correct skewness of the data and approximate a Gaussian distribution. Reported results on predicted CRs are transformed back to the original scale. 
We also analyze the statistical significance of the proposed statistical predictors.  
In the context of linear regression, least absolute shrinkage and selection operator (LASSO) is typically used and acts as an L$^{1}$-regularizer inducing sparsity, hence improving prediction accuracy.  
LASSO regression analysis is expressed as  ${\displaystyle \min _{\beta \in \mathbb {R} ^{p}}\left\{{\frac {1}{N}}\left\|y-X\beta \right\|_{2}^{2}+\lambda \|\beta \|_{1}\right\}}$ with $y$ the response variable (here log(CR)), $X$ the predictors and $\beta$ the regression coefficients to be estimated.  
In practice, $\lambda$ is estimated for each dataset in a cross-validated setup.  
We use the {\rm R}-package {\rm glmnet} to perform the LASSO analysis. 
For spline regressions, a double-penalty method has been proposed in \cite{MARRA2011gam}; it is implemented in the {\rm R}-package {\rm mgcv} and returns a p-value indicating the statistical significance of each predictor. However, it does not provide a quantitative contribution of each predictor. 
Consequently, we discuss predictor contribution for the linear regression model only. 
For both models, statistical predictors are standardized by removing their mean and dividing by their standard deviation.  
This approach enables us to directly compare their relative importance to CRs via the estimated regression coefficients.  
We have an insufficient quantity of data to consider other methods such as random forest or neural networks, as in \cite{moon2022prediction}.

\subsection{Prediction setup and quality evaluation metrics} \label{sec:prediction}
As discussed earlier, only CRs below 100 are considered in the study, since larger ones are rarely achieved in practice due to the user's quality constraints. 
Eq.~\ref{eq:linreg} and~\ref{eq:splinereg} are fitted separately on each set of 2D slices for each field of the datasets and fitted separately for each compressor's CRs.  
To ensure that models are not  overfitted and to reduce biases by randomizing the validation steps, we perform a $k$-fold cross-validation with $k$ varying between 8 and 10 depending on the data size. The data \revise{are randomly} split into $k$ subsets (folds); $(k\!-\!1)$ folds are used to train the regression models, and the remaining fold is used to test \revise{and evaluate} the model in prediction. 
In  cross-validation, training and testing folds are permuted $k$ times; this procedure provides a systematic and unbiased way to assess  out-of-sample prediction error \cite{hastie2009elements}. 
For each \revise{of the $k$ testing folds}, we assess the prediction quality by computing several evaluation metrics between predicted CRs and true CRs: the linear correlation and median absolute percentage error (MedAPE) with the absolute percentage error defined as $\displaystyle \mbox{APE(true, pred)}\! = \! 100\!\times\!\frac{|CR_{true} - CR_{pred}|}{|CR_{true}|}$ and interpretable in $\%$.
The MedAPE offers robust estimate of the accuracy not effected by extremely accurate or inaccurate estimates.
To complement the MedAPE, we report the correlation strength, which is not robust to outliers for several datasets in Table~\ref{tab:5comp_4data_prediction_metrics} suggesting our approach doesn't feature extreme outliers (contra Table~\ref{tab:performance_block}).
We provide confidence intervals based on $10\%$ and $90\%$ quantiles and computed over the $k$ folds for the MedAPE. 
For ease of reading, confidence intervals for the correlation are not shown but available from our scripts. 
\revise{We summarize this in Algorithm~\ref{alg:medape}}

\revise{
\begin{algorithm}
   \caption{Prediction Error Evaluation and Quantification Procedure} 
   \label{alg:medape}
   \begin{algorithmic}
    \REQUIRE D: Dataset, e: error\_bound
    \STATE medape $\gets$ []
    \FOR{train, test $\in$ kfold(D)}
        \STATE true\_cr $\gets$ [], metrics $\gets$ []
        \FOR{d $\in$ train}
            \STATE true\_cr.append(size(compress(d,e))
            \STATE metrics.append([svd(d), qent(d,e)])
        \ENDFOR
        \STATE model $\gets$ train(true\_cr, metrics)
        \STATE ape $\gets$ []
        \FOR{d $\in$ test}
            \STATE true\_cr $\gets$ size(compress(d,e))
            \STATE metrics $\gets$ [svd(d), qent(d,e)]
            \STATE pred\_cr $\gets$ predict(model, metrics)
            \STATE ape.append(100 (true\_cr - pred\_cr) / true\_cr)
        \ENDFOR
        \STATE medape.append(median(ape))
    \ENDFOR
    \RETURN quantile(medape, [0.1, 0.5, 0.9])
   \end{algorithmic}
\end{algorithm}
}

\section{Prediction accuracy results} \label{sec:results}

In this section we discuss results of CR prediction using our method. 
We focus on prediction results from the spline regression Eq.~\eqref{eq:splinereg}, except for Sect. \ref{sec:reg_coeff}, since it is more flexible and provides, on average, higher prediction accuracy. 
For Gaussian samples,  models \eqref{eq:linreg} and \eqref{eq:splinereg} provide similar results. 
For challenging datasets such as the Miranda density, however, the model \eqref{eq:linreg} can lead to a $25\%$ less accurate prediction compared with \eqref{eq:splinereg}. 
First, we discuss prediction results for Gaussian samples as a proof of concept on several compressors (Sect. \ref{sec:res_gaussian}). 
We highlight very competitive CR prediction accuracy that is nonetheless challenged by the increasing complexity of the samples.  
We then explore three different compressor-prediction schemes of SZ3 that are run exclusively (Sect. \ref{sec:sz_modes}); SZ2 typically uses several of them dynamically during a single compression. 
This enables us to  further explore the impacts of the statistical predictors on SZ's different compressor-prediction schemes as well as compare SZ2 with SZ3 and exclusive compressor-prediction schemes.  
In section \ref{sec:res_more}, we provide CR prediction results for various studied datasets and compressors highlighting the robustness of the choice of statistical predictors and our prediction method. 
We discuss results in order to open the discussion toward bridging the CR prediction across error bounds and compressors for single-scale Gaussian samples. 
We compare our method to existing ones on Section \ref{sec:comparison}. 
Finally, Section \ref{sec:results3d} provides results extended to 3D datasets. 
\revise{All the CR prediction results presented in the following are computed in the cross-validation setting described in Sect. \ref{sec:prediction}. }

\begin{figure*}
\centering
\includegraphics[width=.7\textwidth]{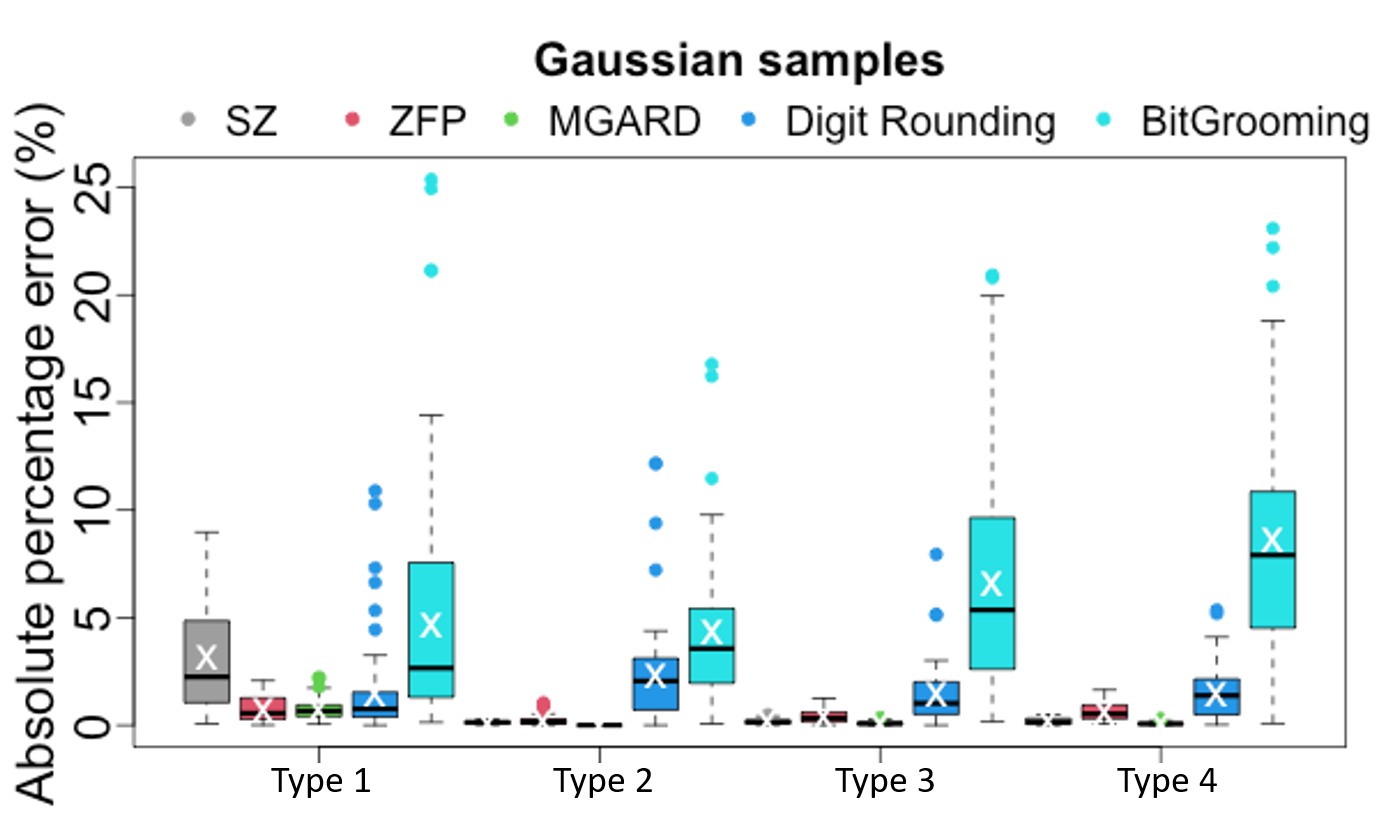}\\[-10pt]
    \caption{Distribution of absolute percentage  error {\small $\frac{|CR_{true} - CR_{pred}|}{|CR_{true}|}$} for the 4 types of Gaussian samples. CR predictions are shown for SZ (grey), ZFP (red), MGARD (green), Digit Rounding (blue), and Bit Grooming (turquoise). White crosses represent the mean of absolute percentage error per box, black lines represent the median, and box outlines represent the 25th (Q1) and 75th (Q3) quantiles. Upper and lower whiskers are min(max(x),Q3+1.5(Q3-Q1)) and max(min(x),Q1-1.5*(Q3-Q1)) with x: data of interest;  typically more than 95\% of the data is contained within the whiskers. Maximum errors are either the upper whiskers or the most up outlier. } 
    \label{fig:prederror_gaussian}
\end{figure*}

\subsection{CR predictions for different types of Gaussian samples} \label{sec:res_gaussian}

We begin our evaluation with a consideration of predictions of CR on Gaussian samples.
Because we can parameterize the Gaussian samples as described in Section~\ref{sec:gaussian} we can use them to evaluate the response of our model to particular types and strengths of correlation in the data to be compressed.

Figure  \ref{fig:prederror_gaussian}  gathers the distribution of absolute  percentage prediction (out-of-sample) error of CR derived following the cross-validation procedure defined in Sect.~\ref{sec:prediction} for various Gaussian samples.  
Since no user specification is available for this data, we choose an absolute error bound of $10^{-3}$, a common error bound of   users, which provides realistic CRs while appropriate to the value range of the samples. 
Overall, the prediction error is very low on the synthetic benchmark data, providing a proof of concept of our method. 
We notice a slight decreasing trend in the prediction accuracy of the CR (although still  competitive) from type 2 to type 4, as the complexity of the spatial heterogeneity increases from scalar  to spatial weights and from fixed to random correlation ranges. 
Gaussian samples of type 4 are the most challenging because correlation scales are picked randomly and aggregated with spatial weights creating samples with strong spatial heterogeneity that may not be encountered in most continuous scientific simulations. 
These samples challenge the chosen statistical predictors and highlight the need to account for heterogeneous multiscale information, as pointed out by  \cite{krasowska2021exploring}.
SZ shows higher errors for samples of Type 1 than for the other sample types, this is due to the fact that  CRs observed in Type 1 samples have a much wider range than the other ones, hence larger prediction errors. 
Rounding-based compressors Digit Rounding and Bit Grooming show the most prediction error (although 75\% of errors is below 10\%) likely due to the fact that  these  compressors  do not leverage spatial structures. 
%
%

%
%
\noindent\fbox{%
    \parbox{0.97\linewidth}{%
        \textbf{Key findings}: CR predictions show competitive  accuracy  (maximum 8\% of MedAPE) on benchmark Gaussian samples and robustness as sample complexity increases.  
    }%
}

\subsection{CR prediction for SZ's compressor-prediction schemes }\label{sec:sz_modes}
\begin{figure}
    \centering
    \includegraphics[scale=.3]{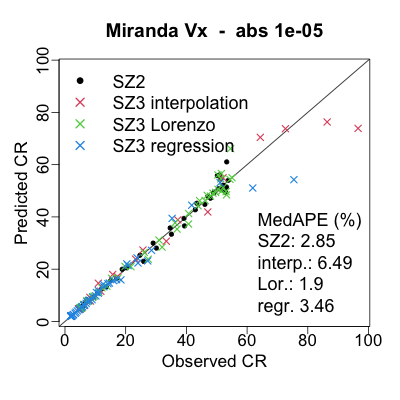}
    \hspace{-.25cm}
      \includegraphics[scale=.3]{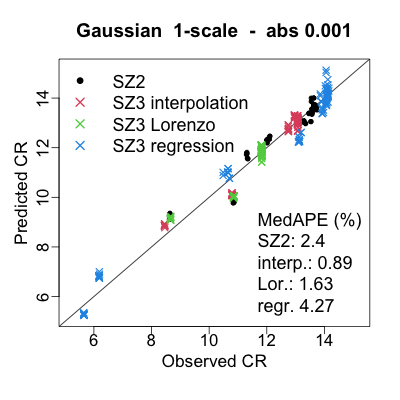} \\
      \includegraphics[scale=.3]{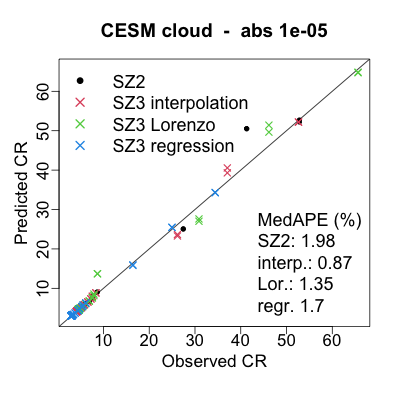} \hspace{-.25cm}
       \includegraphics[scale=.3]{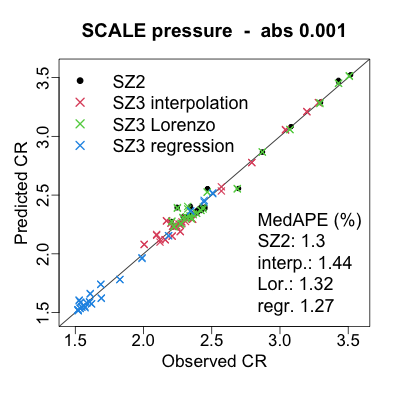}

     \caption{True (x-axis) and out-of-sample predicted (y-axis) CR for  SZ2 with dynamic selection of compressor-prediction scheme (black dot) and 3 prediction schemes run individually in SZ3: interpolation (red), Lorenzo (green), and regression (blue). CR prediction performed with spline regressions. Results shown for an 8-fold cross-validation procedure (all predicted folds and corresponding truth are shown) in absolute error bound for individual fields of Miranda ($10^{-5}$), Type 1 Gaussian samples ($10^{-3}$), CESM-cloud $10^{-5}$ and \revise{SCALE-LetKF} pressure ($10^{-3}$). Median absolute percentage error of the CR prediction are plotted in each panel for each compressor.   }
    \label{fig:sz_3modes}
\end{figure}

\begin{table*}
\caption{Correlations and  MedAPE (\%) for CR prediction with spline regression in 8-fold cross-validation; 10\% and 90\% quantiles of  MedAPE are reported in parentheses.} 
    \centering
    \begin{tabular}{l|c|c|c|c}
     & \multicolumn{2}{c|}{Miranda Vx abs $10^{-5}$} & \multicolumn{2}{c}{Miranda De abs $10^{-5}$} \\
       & Corr. & MedAPE  (\%) & Corr. & MedAPE (\%) \\
      \hline
      SZ2   & 0.998 & 2.8 (0.9,3.8) & 1.0 & 1.3 (0.7,19.8)  \\
      \hline
      ZFP & 0.999 & 1.0 (0.6,1.7) & 0.995 & 2.1 (0.6,6.7) \\
      \hline
     MGARD  & 0.999 & 1.9 (1.8,2.7) & 1.0 & 6 (2.0,10.2) \\
     \hline
     Digit Rounding & 0.995 & 0.2 (0.1,0.2) & 0.953 & 11.9 (7.0,13.1) \\
     \hline 
     \hline      & \multicolumn{2}{c|}{NYX Vx abs=$10^{-2}$} & \multicolumn{2}{c}{SCALE U abs=$10^{-3}$} \\
       & Corr. & MedAPE (\%) & Corr. & MedAPE (\%) \\
      \hline
      SZ2   & 0.176 & 1.0 (0.9,1.0) & 0.710 & 9.7 (7,12.3) \\
      \hline
      ZFP  & 0.75 & 0.9 (0.9,1.0) & 0.162 & 3.1 (2.5,6.6) \\
      \hline
     MGARD   &  0.701 & 0.2 (0.2,0.2) & 0.655 & 3.7 (1.4,5.8) \\
      \hline
     Digit Rounding  & 0.331 & 7.0 (6.2,8.1) & 0.918 & 1.2 (1.0,1.94) \\
     \hline 
     \hline      & \multicolumn{2}{c|}{CESM cloud abs=$10^{-5}$} &  \multicolumn{2}{c}{Hurricane U abs=$10^{-2}$} \\
       & Corr. & MedAPE (\%) & Corr. & MedAPE (\%) \\
      \hline
      SZ2   & 1 & 1.94 (1.3,2.7) & 0.876 & 2.4 (1.3,3.6) \\
      \hline
      ZFP  & 1 & 0.5 (0.2,0.9) & 0.957 & 0.8 (0.5,1.3) \\
      \hline
     MGARD   & 1 & 0.8 (0.3,1.1) & 0.947 & 0.9 (0.4,1.2) \\
      \hline
     Digit Rounding  & 1 & 1.6 (1.1,3.1) & 0.493 & 1.3 (0.6,2.4) \\
    \end{tabular}
    \label{tab:5comp_4data_prediction_metrics}
\end{table*}

In this section, we provide CR-prediction results for SZ3 running with a fixed exclusive compressor-prediction scheme in the compression.
Because we can choose the compressor-predictor scheme in SZ3, we can use SZ3 to consider how robust our method is to variation in a compression scheme and show the robustness of the accuracy of its predictions over the evolution of a compressor.

We consider the three compressor-prediction schemes introduced over time to the SZ series: Lorenzo (v1), regression (v2), and interpolation (v3).  We compare the results with those from SZ2 running with dynamic selection between regression and Lorenzo which is the hardest case to predict accurately. 
This  enables to further understand the response of SZ's various individual compressor-prediction scheme to statistical predictors and goes further than existing  studies \cite{qinEstimatingLossyCompressibility2020, krasowska2021exploring} that focus on SZ1.4, which is missing the regression and interpolation schemes. These two compressor-prediction schemes have been critical in improving compression ratios at high-compression use cases \cite{liangErrorControlledLossyCompression2018, zhaoOptimizingErrorBoundedLossy2021}.

In Fig. \ref{fig:sz_3modes}, we show scatterplots of true and (out-of-sample) predicted CRs that are computed in cross-validation (see Sect.~\ref{sec:prediction}). We expect both data to match along the first diagonal.
Overall, SZ2 and SZ3 with fixed exclusive compressor-prediction schemes provide a very good matching between true and predicted CRs and very low median percentage error, as displayed on each panel. 
This highlights  the robustness of our CR-prediction method to complex compression schemes as SZ2 runs with a dynamic selection of prediction schemes and its CR predictions remain as accurate as predictions for SZ3 with single-prediction mode. 
We note that different compressor-prediction schemes tend to provide different values of CRs hence leading to different quality of CR prediction.
However, we notice that across datasets SZ3 achieves only marginally higher CR than does SZ2 with dynamic selection of prediction schemes (with accurate associated prediction). 
This highlights the optimality of  SZ2's dynamic selection of prediction modes. 
Furthermore, we  observe that  SZ3 with exclusive regression-prediction scheme tends to generate lower CRs (although well predicted by our technique) than the other exclusive prediction schemes and lower than SZ2 with dynamic selection of prediction schemes. 
This is well captured by SZ2 with dynamic selection of compressor-prediction schemes. 
Indeed, for Miranda fields a maximum of $9.5\%$ of blocks per 2D slice are predicted with regression, for CESM-cloud the median percentage is $5.7\%$,for the pressure field of \revise{SCALE-LetKF}, this maximum percentage per 2D slice drops to $0.05\%$ and for Gaussian samples, the median percentage of use of regression is $93.5\%$. 
This suggests that our CR prediction technique could be used to complement the selection steps performed in SZ2 when used with dynamic selection and serve as a proxy for selecting prediction schemes. This is left for future work. 
Our proposed CR-prediction technique performs equally well from high CRs observed on Miranda velocity-x field to low CRs observed on the \revise{SCALE-LetKF} pressure field. 
In practice compression on the \revise{SCALE-LetKF} pressure field would be performed with relative error bound because of the very high values of the fields. 
However, we keep the absolute error bound for consistency of the study. 

The importance of the different statistical predictors (quantized entropy, $\frac{\mbox{SVD trunc}}{\sigma}$ and their interaction) is analyzed via a LASSO analysis for the linear regression (that provides an overall trend approximation) as discussed in Sect. \ref{sec:methods}. 
SZ2 and SZ3 with its fixed exclusive compressor-prediction schemes exhibit similar trends in the statistical predictor importance. 
Predictor importance varies significantly across datasets; however, in each case both spatial correlation and entropy information are significant, either as separate  predictors or through their interaction term. 
This highlights the flexibility of the proposed CR prediction model and the choice of statistical predictors. 
\noindent\fbox{%
    \parbox{0.97\linewidth}{%
        \textbf{Key findings}: CR-prediction methods show similar prediction accuracy (with less than 5\% difference) for the complex SZ2 with dynamic selection of compressor-prediction scheme and SZ3 with fixed compressor-prediction scheme, highlighting the versatility of the CR-prediction method.  }
}

\subsection{CR prediction for additional compressors and datasets}\label{sec:res_more}
\begin{figure*}
    \centering
    \includegraphics[width=.7\textwidth]{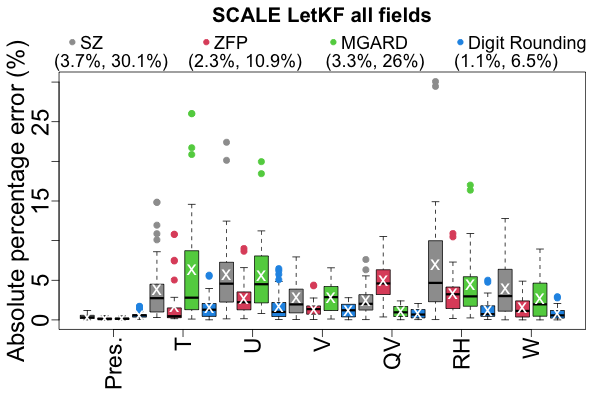}
    \caption{Distribution of absolute prediction error {$\frac{|CR_{true} - CR_{pred}|}{|CR_{true}|}$} for fields of \revise{SCALE-LetKF}. CR predictions shown for SZ (grey), ZFP (red), MGARD (green), and Digit Rounding (blue). Numbers in parentheses correspond to the average and maximum error over all shown fields. 
    White crosses represent the mean of absolute percentage error per box, black lines reprresent the median, and the box outlines represent the 25th (Q1) and 75th (Q3) quantiles. Upper and lower whiskers are min(max(x),Q3+1.5(Q3-Q1)) and max(min(x),Q1-1.5*(Q3-Q1)) (x: data of interest). Maximum errors are either the upper whiskers or the most up outlier. }  
    \label{fig:prederror_allfields}
\end{figure*}

\begin{figure}
    \centering
    \includegraphics[scale=.65]{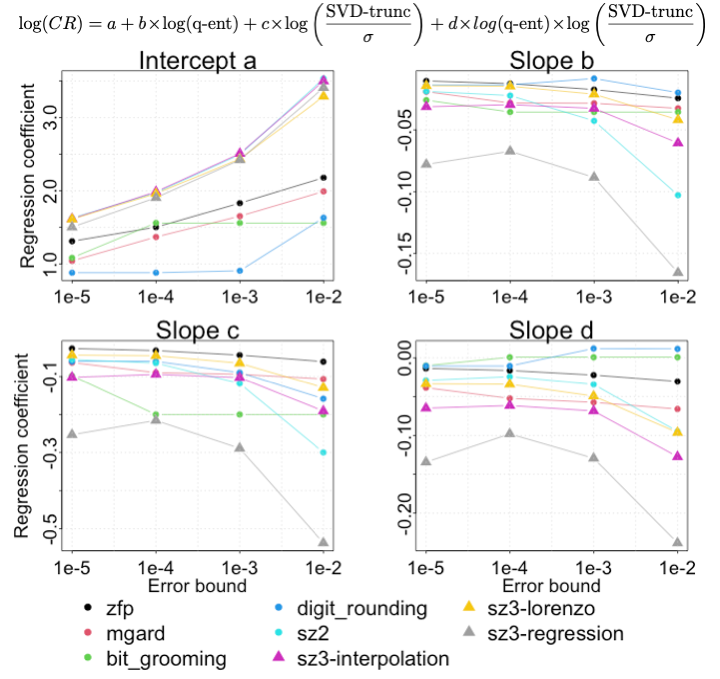}\\[-10pt]
    \caption{Coefficients (y-axis) of the linear regression  Eq. \eqref{eq:linreg} fitted on single-scale Gaussian samples for 8 compressors (different colors). Regressions are fitted independently for each compressor and each error bound (x-axis), lines between points only help the reading and do not represent functional fits. Coefficients can be interpreted relatively as predictors have been normalized.  }
    \vspace{-0.20cm}
    \label{fig:regression_coefficient}
\end{figure}

In this section, we broaden our examination to assess the accuracy of our approach for a variety of different datasets and compressors to demonstrate the robust viability and accuracy of our approach.

Figure \ref{fig:scatterplot_prediction} shows examples of  CR predictions through matching of out-of-sample predictions with true CR. 
The linearity of the plot shows that we accurately predict the compression ratio for variety of compressors using a variety of underlying principles.

In Table \ref{tab:5comp_4data_prediction_metrics}, we collect metrics discussed in Sect.~\ref{sec:prediction} for compressors with different principles: prediction-based (SZ2), transformation-based (ZFP), multigrid-based (MGARD), and rounding-based (Digit Rounding) with additional datasets. 
Miranda's density and \revise{SCALE-LetKF}'s U field show the least prediction accuracy because of their spatial heterogeneity. 
\revise{SCALE-LetKF}'s U field presents a lot of spatial heterogeneity with small-scale features blended with large-scale ones; see Fig.~\ref{fig:gaussian_sample}. 
On the other hand, Miranda's density exhibits a ``polka-dot-like'' structure (see Fig. 5 of \cite{zhaoSdrbench2020}). This is highlighted in the predictor contributions where quantized entropy and its interaction with SVD truncation are the most significant statistical predictors for SZ2, ZFP,  and MGARD. 

Figure \ref{fig:prederror_allfields} shows further explorations of the prediction error distributions for several fields of \revise{SCALE-LetKF} (fields providing enough datapoints to fit regression robustly). Median and mean absolute percentage prediction errors are consistent with previous results and consistent with averages over all predicted fields.  Most maximum errors remain acceptable and competitive.
All maximum error predictions are less than $30\%$ for \revise{SCALE-LetKF}.  Results are consistent with other datasets.

Additionally, we discuss the sensitivity of the CR prediction accuracy to the training dataset size. For instance, reducing the training set  for \revise{SCALE-LetKF} U from 70\% to 20\% of the full dataset leads to a decreased MedAPE by 34\% for SZ, 39\% for ZFP, 79\% for MGARD, and 26\% for Digit Rounding. However, results vary depending on how compressors and datasets create homogeneity in the training set. For instance, for Miranda Vx the best MedAPE results are achieved with around 30\% of the entire set as a training set. This holds promises for  3D settings with  fewer available samples.  

\revise{Finally, in Table \ref{tab:sa_lasso}, the importance of the different predictors (quantized entropy, $\frac{\mbox{SVD trunc}}{\sigma}$ and their interaction) is analysed via a LASSO analysis for the linear regression (that provides an overall trend approximation) as discussed in Sect. \ref{sec:methods} and for SZ. 
Predictors importance vary significantly across datasets; however in each case both spatial  correlation and entropy information are significant either as separate  predictors or as an interaction term. 
This highlights the flexibility of the proposed regression model and the choice of predictors.}
\revise{\begin{table}
    \caption{Predictor importance for SZ from 8-fold cross-validation of LASSO regression. Coefficients shown in absolute values to be interpreted as a relative importance to compression ratio prediction. The dot $\cdot$ indicates an insignificant  predictor to CR according to LASSO criteria. }
    \begin{tabular}{p{3.5cm}|c|c|c}
     & q-ent. & $\frac{\mbox{SVD}}{\sigma}$ & q-ent*$\frac{\mbox{SVD}}{\sigma}$ \\
      \hline
     Miranda Vx {\footnotesize abs $10^{-5}$}   &  1.10 & $\cdot$ & 0.48 \\
      \hline 
     CESM cloud {\footnotesize abs $10^{-5}$} & 0.69 & 0.01 & 0.02 \\ 
     \hline    
     Gaussian 1-scale {\footnotesize abs $10^{-3}$} & 0.04 & 0.12 & 0.03 \\
    \hline
   SCALE pressure {\footnotesize abs $10^{-3}$} & 0.01 & 0.12 & 0.02 \\
    \end{tabular}
    \label{tab:sa_lasso}
\end{table}
}
%
%

%

%
%

\noindent\fbox{%
    \parbox{0.97\linewidth}{%
        \textbf{Key findings}: Median percentage error is overall  low (less than 12\%) across compressors and datasets, highlighting the flexibility and accuracy of the method. 
        
    }%
}

\subsection{Toward CR characterization across compressors}\label{sec:reg_coeff}
Understanding the strength of the relationships between our predictors and compression ratios could enable a possible path to providing a model of compress-ability across compressors and enable us to describe which aspects of data are best captured by a particular compressor.
In Figure \ref{fig:regression_coefficient}, we compare coefficients of the regression  Eq. \eqref{eq:linreg} for the 8 compressors and 4 absolute error bounds ($10^{-5}$, $10^{-4}$, $10^{-3}$, $10^{-2}$). Regressions are fitted on Gaussian samples with a single-scale correlation since they serve as a benchmark.  
To ease the reading,  we show results for the linear regression. One can imagine a similar plot for the spline regression Eq. \eqref{eq:splinereg}  with the estimated spline coefficients. 
Regression predictors have been normalized, one can then interpret their relative importance via the estimated coefficients.

Regression predictors have been normalized so estimated coefficients represent their relative importance. 
For most compressors, transitions from one error bound to another are simple and smooth and can be represented by a low-order polynomial function.  
This raises the opportunity of future work to build statistical models to predict CR and account for the error bound as a parameter. Regression coefficients $a$, $b$, $c$, and $d$ would be modeled and fitted as functions of the error bound, enabling CR prediction across error bounds via a statistical model.  
This indicates the varying importance of the predictors with changing error bounds. 
The intercept $a$ naturally increases as the error bound  decreases since  CRs increase.
The coefficient $c$ for most compressors, the importance of the predictor  $\frac{\mbox{SVD-trunc}}{\sigma}$  increases or plateaus as the error bound increases, indicating that for more permissive error bounds, compression relies more on spatial correlations. 
This increasing trend is also observed for the other predictors. 
Moreover, similar trends  are shared by several compressors, providing information to understand compressibility across compressors.   
Digit Rounding and Bit Grooming are almost insensitive to the interaction term between the quantized entropy and $\frac{\mbox{SVD-trunc}}{\sigma}$ and rely the most on the quantized entropy. 
This  is expected since Bit Grooming and Digit Rounding do not account for spatial factors. 
ZFP  and MGARD exhibit similar  linear trends across error bounds, whereas SZ2 with dynamic selection of prediction mode and SZ3 with fixed prediction modes tend to show almost quadratic behaviors across error bounds.  
Note that SZ3-regression and SZ2 show  similar overall trends; the reason is that for this dataset SZ2 has at least a median of 90\% blocks predicted by regression across error bounds. 
Overall, the importance of the quantized entropy (slope $b$) is less as a single predictor than as an interaction term (slope $d$). This corroborates Fig.~\ref{fig:individual_predictor_miranda} showing complementary skills of $\frac{\mbox{SVD-trunc}}{\sigma}$ and the quantized entropy as explanatory variables of the CR. 
\noindent\fbox{%
    \parbox{0.97\linewidth}{%
        \textbf{Key findings}: Compressibility patterns characterized via regression coefficients show consistency depending on compressor type and smooth transitions across error bounds. 
    }%
}

\subsection{CR prediction results for 3D datasets}\label{sec:results3d}
While many scientific datasets are 2D, there are also many 3D scientific datasets, and it is important to also be able to predict compression ratios on these datasets.
In this section, we assess our performance on 3D datasets.
Prediction is performed out-of-sample following the cross-validation procedure described in Sect.~\ref{sec:prediction}.
We now additionally consider TTHRESH which is designed for only for 3D and higher dimensional data.

\begin{figure}
    \centering
    \includegraphics[scale=.3]{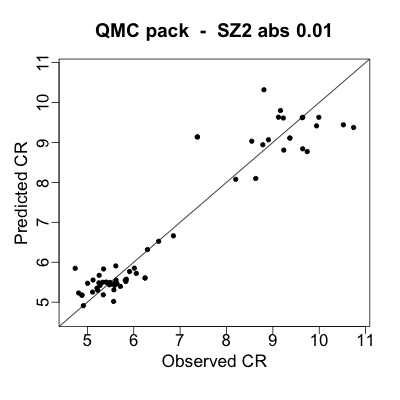}%
 \includegraphics[scale=.3]{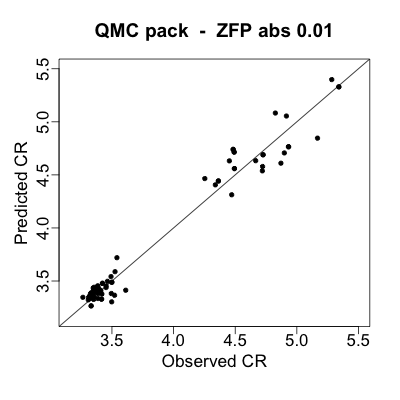} \\
    \caption{Scatterplots of true and out-of-sample predicted CR from spline regressions. Results are shown for 3D events of the QMCPACK dataset for SZ2 (left) and ZFP (right) with $10^{-2}$ absolute error bound. These plots show a MedAPE of 4.1 and a MaxAPE of 24} 
    \label{fig:res_QMCPACK}
\end{figure}

We present an exemplar in Figure \ref{fig:res_QMCPACK} which shows the quality of CR statistical predictions  for the 3D dataset QMCPACK.
While the spread of the accuracy appears higher than what is observed in Figure~\ref{fig:scatterplot_prediction}, the MedAPE is still quite small at 4.1\% which is consistent with the results presented elsewhere in the paper.

Broadening our consideration, the overall quality of CR predictions remains competitive across most compressors as observed on Table \ref{tab:performance_3d} in particular for SZ2, ZFP, and MGARD. 
While the CRs for TTHRESH are the most challenging to predict with our method but still have much less error than other fast methods had for other compressors (e.g., Table~\ref{tab:performance_block}) which incur MedAPE errors at or above 76\% representing a $3\times$ improvement in MedAPE for TTHRESH. 
\begin{table}
\tabcolsep=0.1cm
\caption{Prediction accuracy metrics for SZ2, ZFP, MGARD, Bit Grroming, and TTHRESH's CR. MedAPE (\%) with 10\% and 90\% quantiles are reported for QMCPACK data.  }
   \centering
   \begin{tabular}{l|rr}
     Compressor & MedAPE & (10\% APE, 90\% APE)  \\
     \hline
     SZ2 & 4.5 & (3.2, 5.7)  \\
     ZFP & 1.7 & (1.3, 3.5) \\
     MGARD & 0.6 & (0.4, 1.3) \\
     BitGrooming & 7.4 & (5.0, 9.3) \\ 
     TTHRESH & 24.8 & (15.7, 27.7)   \\
    \end{tabular}
    \label{tab:performance_3d}
\end{table}
Similar associations and complementarity between 3D statistical predictors and CR are observed on other datasets such as Miranda, but we do not have enough samples to fit our regression model.
Future work will focus on sampling 3D datasets to create samples in order to fit statistical prediction models.

\noindent\fbox{%
    \parbox{0.97\linewidth}{%
        \textbf{Key findings}: Extension to 3D data of CR statistical predictions  remains competitive for SZ, ZFP, MGARD, and Bit Grooming. Predictions for TTHRESH are less accurate , but still substantially better than using block sampling estimation. 
    }%
}

\subsection{Accuracy comparison to prior work}\label{sec:comparison}


In this section, we compare against the state-of-the-art estimation methods. 
For more information on these methods see Section~\ref{sec:literature}.
We present results here on SZ2 which is the hardest compressor to predict correctly, but results on other compressors are similar.

\begin{table}
\tabcolsep=0.1cm
\caption{Prediction accuracy metrics for prior work on SZ2. MedAPE (\%) with 10\% and 90\% quantiles are reported for Miranda Vx  and CESM CLOUD.  }
\centering
   \begin{tabular}{p{2.3cm}|p{2.5cm}|p{2.5cm} }
   \footnotesize
      & Miranda Vx & CESM CLOUD  \\
      \hline
    Our method & 2.8 (0.9, 3.8) &  1.9 (1.3, 2.7)  \\
      \hline
    OptZConfig \cite{underwoodOptZConfigEfficientParallel} &  28 (17, 121) & 26 (12, 58) \\
     \hline
    Block sampling \cite{taoOptimizingLossyCompression2019} & 90 (82, 93) & 76 (41, 82) \\
    \hline
    Method from \cite{luUnderstandingModelingLossy2018} & 193 (157, 276) &  1398 (713, 1570)  \\
    \end{tabular}
    \label{tab:performance_block}
\end{table}
 Table \ref{tab:performance_block} provides SZ2 CR-prediction accuracy for several fields via block sampling \cite{liangImprovingPerformanceData2019, taoOptimizingLossyCompression2019}, via the compressor-specific method from \cite{luUnderstandingModelingLossy2018}, OptZConfig \cite{underwoodOptZConfigEfficientParallel}, and our approach. 
Predictions from each method are implemented from their codes.   

Results from these three other methods are not as competitive as our method; in particular, block sampling systematically underestimates CRs whereas \cite{luUnderstandingModelingLossy2018}'s method systematically overestimates them. 
It is important to note that the next most accurate approach by \cite{underwoodOptZConfigEfficientParallel} is substantially slower -- we explore the performance aspect more fully in Section~\ref{sec:perf}.
Note that \cite{luUnderstandingModelingLossy2018}'s method was designed mainly for 1D datasets; and, as discussed earlier, block-sampling techniques were designed to  predict only whether  SZ  or  ZFP would provide higher CR and leverage the compressor block size.

\noindent\fbox{%
    \parbox{0.97\linewidth}{%
        \textbf{Key findings}: Our approach is at least an order of magnitude more accurate than prior approaches with at least a $10\times$ improvement over OptZConfig, and  at least $32\times$ than block sampling, and $69\times$ than \cite{luUnderstandingModelingLossy2018}.
    }%
}

\section{Performance analysis of our approach vs prior work}\label{sec:perf}
\begin{figure*}
     \centering
     \includegraphics[width=\textwidth]{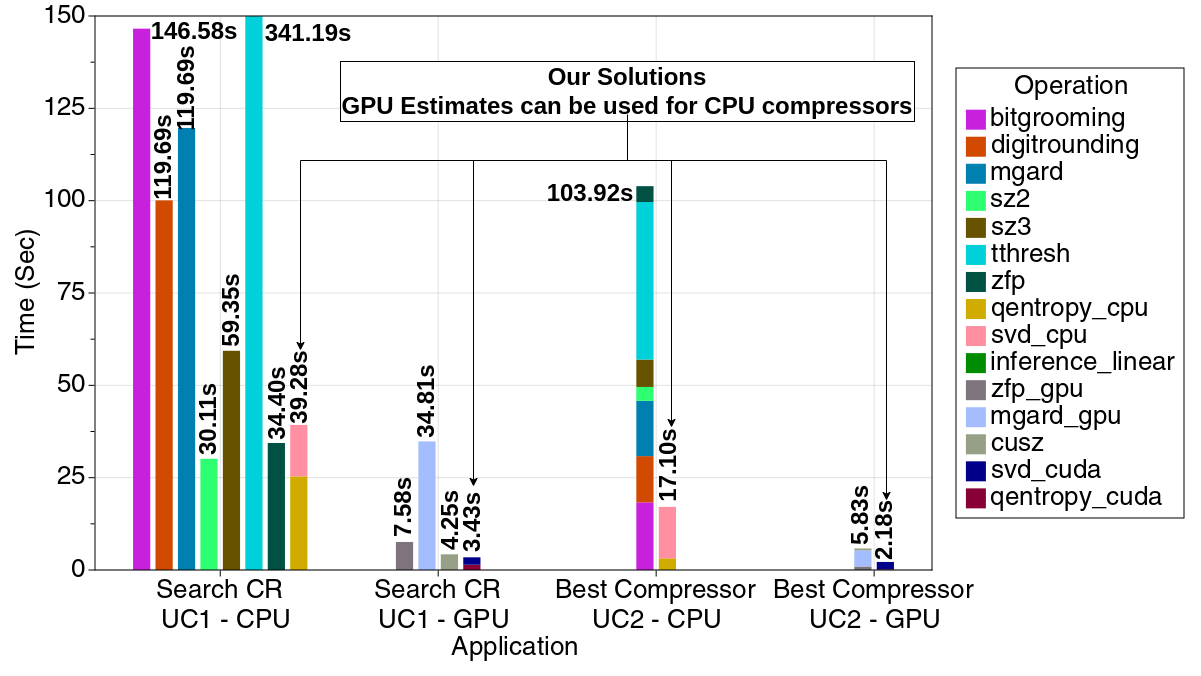}
      \caption{Runtime for use case 1 and use case 2 on \revise{SCALE-LetKF} V:  Our GPU-based approach is the fastest, with 3.4 s for use case 1 (UC1) and 2.0 s for use case 2 (UC2).  Compressors take more time in use case 1 because they are iteratively executed in OptZConfig~\cite{underwoodOptZConfigEfficientParallel}. Even without access to a GPU, our CPU-based implementation is faster than all CPU compressors for use case 1 except SZ2 and ZFP at 39.2s. TTHRESH is especially slow in use case 1, taking 341 seconds.  There are 12 implementations of use case 1 and 4 of use case 2}
    \label{fig:my_label}
\end{figure*}
An accurate method to predict compress-ability is limited if it cannot be used to accelerate applications using compression.
This section investigates the performance and feasibility of using our approach for the two use cases from the introduction.
Use case 1: Determine a configuration of a compressor that achieves a specified CR \cite{underwoodFRaZGenericHighFidelity2020}.
Use case 2: Determine which of a set of compressors achieves the best CR \cite{taoOptimizingLossyCompression2019}.

We include the GPU versions of SZ, ZFP, and MGARD here for performance comparison but omit their quality results because of lack of space and in implementation issues in some of the GPU compressors (i.e. segfaults).  Only ZFP's GPU implementation ran without crashes on our entire testing sets, and ZFP's GPU implementation only supports fixed rate, which is by definition trivial to estimate CR but has lower quality than error bounded modes.
We specifically  use \revise{SCALE-LetKF} V data for our test because it has the largest buffers in our dataset and thus a worse case for our approach due to the runtime complexity of the SVD.

For use case 1, we compare against OptZConfig \cite{underwoodOptZConfigEfficientParallel} in a warm-start case \cite{underwoodFRaZGenericHighFidelity2020}.
OptZConfig is able to target a specific compression ratio by iteratively executing the compressor and using black-box global optimization to configure the compressor appropriately to achieve the specified target.
By warm start, we assume that we already have a trained model for both OptZConfig and our approach for other data from this field.
In contrast to OptZConfig, we present our approach implemented on both CPUs and GPUs.
In both implementations, we still use the optimization approach from OptZConfig; but instead of using the actual compressor, we use our statistical approach to estimate the CR.
Since the SVD or HOSVD portion is independent of the error bound, we  execute this code only once; however, the quantized entropy and our inference code is run for each error bound.
We  report the wall-clock time for each segment of the process;  sum is the total time taken.

For use case 2, we compare against running each compressor once to determine which has the greatest CR \cite{taoOptimizingLossyCompression2019}.
We still assume a warm start where we have trained on other data from the same field.
In this case, we  need to run both the SVD/HOSVD and quantized entropy only once and our inference code once per compressor we compare against.
We  report the wall-clock time for each segment of the process;  sum is the total time taken.

All compressors were compiled with default flags and latest versions from Spack.
Our implementations for both versions of our metrics are parallel and written in Julia, whereas the inference code is written in R.
The implementation of the quantized entropy is straightforward from its definition with a parallel reduction.
The implementation of the SVD/HOSVD is also reasonably straightforward, with the bulk of the parallelism expressed via a parallel LAPACK library: MKL for the CPU, cuSOLVER for the GPU.
Note that our time to estimate does not vary based on compressor and that we achieve equivalent quality estimates from both our method and the baselines so we do not present them here.

For use case 1, our GPU-based approach achieves substantial speedups compared with  executing the compressors and our approach on the CPU, with speedups between $10\times$ and $1.25\times$ for GPU compressors, $100.0\times$ to $8.9\times$ over the CPU compressors, and $11.2\times$ over the CPU parallel version of our code.
Even the slower CPU version outperforms the compressors with OptZConfig in 5 of 7 CPU compressors.
For use case 2, our GPU-based approach achieves a $50\times$ speedup with all compressors and still a $21.5\times$ speedup excluding the two slowest compressors and a $4\times$ speedup with just the fastest two.
The CPU-based approach is still able to achieve a $1.3\times$ speedup when 4 or more compressors are used and $4.9\times$ speedup with all compressors.
In both cases we observe similar prediction accuracy as reported in  earlier sections.

\section{Conclusion} \label{sec:conclusion}

We propose an accurate statistical method to predict CR of lossy compressors using statistical properties of data that is agnostic to compressor internals. 
Compressor-free statistical predictors are based on spatial correlation, entropy, and lossyness via  SVD-based and entropy-based predictors which enables speedups compared to the existing literature.  
CR prediction accuracy is highly competitive in terms of prediction accuracy, with very low percentage error for the studied compressors.   
Our method demonstrates robustness to a variety of compressors which rely on differing operating principles.
We even show accurate CR predictions for SZ2, highlighting the robustness of the method to the dynamic selection of compressor-prediction schemes in SZ2. 
Future work should focus further on the generalizability of the method to reduce dependence on samples, types of bounds, and compressors principles. 

This method enables speedups in both automated tuning of compressors to find a desired compression ratio (use case 1, \cite{underwoodFRaZGenericHighFidelity2020}) as well as quickly determining which compressor out of a groups of compressors will achieve the greatest CR (use case 2, \cite{taoOptimizingLossyCompression2019}).
Both use cases have been used to accelerate IO for parallel computing.

\section*{Acknowledgments}
The first two authors contributed equally to the paper. This material is based upon work supported in part by the Exascale Computing Project (17-SC-20-SC) of the U.S. Department of Energy (DOE), and by DOE’s Office of Science, under contract DE-AC02-06CH11357. This material is also based upon work supported by the National Science Foundation under Grant No. SHF-1910197, SHF-1943114, OAC-2003709 and OAC-2104023. We gratefully acknowledge the computing resources provided on Bebop, a high-performance computing cluster operated by the Laboratory Computing Resource Center at Argonne National Laboratory. Clemson University is acknowledged for generous allotment of compute time on the Palmetto cluster. 

\bibliography{compressability-citations}
\bibliographystyle{SageH}

\end{document}